\documentclass[12pt]{article}

\usepackage{amssymb}
\include{epsf}

\newcommand{\A}{{\sf A}}

\newcommand{\K}{{\rm K}}
\newcommand{\HH}{{\rm H}}

\newcommand{\ee}[1]{{\rm e}^{#1}}
\newcommand{\ii}{{\rm i}}

\def\Dirac{{D\!\!\!\!/\,}} % Dirac operator

\newcommand{\eq}{\begin{equation}}
\newcommand{\eqend}{\end{equation}}
\newcommand{\eqa}{\begin{eqnarray}}
\newcommand{\nonueqa}{\begin{eqnarray*}}
\newcommand{\eqaend}{\end{eqnarray}}
\newcommand{\nonueqaend}{\end{eqnarray*}}

\newcommand{\bma}[1]{\begin{array}{#1}}
\newcommand{\ema}{\end{array}}
\newcommand{\bc}{\begin{center}}
\newcommand{\ec}{\end{center}}

\newcommand{\newsection}[1]
{\vspace{5mm}
\pagebreak[3]
\addtocounter{section}{1}
\setcounter{equation}{0}
\setcounter{subsection}{0}
\begin{flushleft}
{\large\bf \thesection. #1}
\end{flushleft}
\nopagebreak
\medskip
\nopagebreak}

\newcommand{\newsubsection}[1]{
 \vspace{5mm}
\pagebreak[3]
\addtocounter{subsection}{1}
\noindent{ \bf \thesubsection. #1}
\nopagebreak
\vspace{2mm}
\nopagebreak}

\def\appendix#1{\addtocounter{section}{1}\setcounter{equation}{0}
\renewcommand{\thesection}{\Alph{section}}
\begin{flushleft}
{\large\bf \thesection. #1}
\end{flushleft}
\nopagebreak
\medskip
\nopagebreak}

\newcommand{\complex}{{\bb C}} %% complex numbers
\newcommand{\zed}{{\bb Z}} %% integers
\newcommand{\real}{{\bb R}} %% real numbers
\newcommand{\sphere}{{\bb S}} %% sphere
\newcommand{\spheres}{{\bbs S}} %% small sphere
\newcommand{\spheress}{{\bbss S}} %% smaller sphere
\newcommand{\ball}{{\bb B}} %% ball
\newcommand{\balls}{{\bbs B}} %% small ball
\newcommand{\zeds}{{\bbs Z}} %% small integers
\newcommand{\rat}{{\bb Q}} %% rational numbers
\newcommand{\id}{{1\!\!1}} %% identity matrix
\def\Dirac{{D\!\!\!\!/\,}} % Dirac operator
\newcommand{\ch}{{\rm ch}} %% Chern character
\newcommand{\SlashN}{\nabla\!\!\!\!/\,}
\newif\ifold             \oldtrue

\font\mybb=msbm10 at 12pt
\def\bb#1{\hbox{\mybb#1}}
\font\mybbs=msbm10 at 9pt
\def\bbs#1{\hbox{\mybbs#1}}
\font\mybbss=msbm10 at 7pt
\def\bbss#1{\hbox{\mybbss#1}}

\def\nn{\nonumber}

\newcommand{\Tr}[1]{\:{\rm Tr}\,#1}

\def\e{{\,\rm e}\,}

\def\be{\begin{equation}}
\def\ee{\end{equation}}
\def\bea{\begin{eqnarray}}
\def\eea{\end{eqnarray}}
\def\bd{\begin{displaymath}}
\def\ed{\end{displaymath}}

\def\dd{{\rm d}}

\def\ii{{\,{\rm i}\,}}

\newcommand{\beq}{\begin{eqnarray}}
\newcommand{\eeq}{\end{eqnarray}}

\makeatletter
\newdimen\normalarrayskip              % skip between lines
\newdimen\minarrayskip                 % minimal skip between lines
\normalarrayskip\baselineskip
\minarrayskip\jot
\newif\ifold             \oldtrue            
\def\arraymode{\ifold\relax\else\displaystyle\fi} % mode of array entries
\def\@arrayskip{\ifold\baselineskip\z@\lineskip\z@
     \else
     \baselineskip\minarrayskip\lineskip2\minarrayskip\fi}
\def\@arrayclassz{\ifcase \@lastchclass \@acolampacol \or
\@ampacol \or \or \or \@addamp \or
   \@acolampacol \or \@firstampfalse \@acol \fi
\edef\@preamble{\@preamble
  \ifcase \@chnum
     \hfil$\relax\arraymode\@sharp$\hfil
     \or $\relax\arraymode\@sharp$\hfil
     \or \hfil$\relax\arraymode\@sharp$\fi}}
\def\@array[#1]#2{\setbox\@arstrutbox=\hbox{\vrule
     height\arraystretch \ht\strutbox
     depth\arraystretch \dp\strutbox
     width\z@}\@mkpream{#2}\edef\@preamble{\halign \noexpand\@halignto
\bgroup \tabskip\z@ \@arstrut \@preamble \tabskip\z@ \cr}%
\let\@startpbox\@@startpbox \let\@endpbox\@@endpbox
  \if #1t\vtop \else \if#1b\vbox \else \vcenter \fi\fi
  \bgroup \let\par\relax
  \let\@sharp##\let\protect\relax
  \@arrayskip\@preamble}
\makeatother

\begin{document}

\begin{flushright}

\baselineskip=12pt

HWM--02--30\\
EMPG--02--19\\
hep--th/0209210\\
\hfill{ }\\
September 2002
\end{flushright}

\begin{center}

{\large\bf D-BRANES, TACHYONS AND
  K-HOMOLOGY\footnote{\baselineskip=12pt Based on invited lecture
  given at the Workshop on Algebraic Geometry and Physics ``K-Theory,
  Derived Categories and Strings'', University of Genoa, Genoa, Italy,
  June 18--21 2002.}}

\baselineskip=12pt

\vspace{.5cm}

{\bf Richard J. Szabo}
\\[3mm]
{\it Department of Mathematics, Heriot-Watt University \\ Riccarton,
  Edinburgh EH14 4AS, U.K.}
\\{\tt R.J.Szabo@ma.hw.ac.uk}
\\[10mm]

\end{center}

\begin{center}
\begin{minipage}{15cm}
\small

\baselineskip=12pt

We present an overview of the ways in which D-brane charges are classified in
terms of K-theory, emphasizing the natural physical interpretations of
a homological classification within a topological setting.

\end{minipage}
\end{center}

\baselineskip=14pt

\newsection{Introduction \label{Intro}}

In this lecture we will discuss one of the most interesting recent
developments in mathematical physics, the classification of D-brane
charges in string theory using K-theoretic
methods~\cite{MinMoore}--\cite{MooreWitten1}. Our presentation will be in
the context of explaining how certain mathematical structures find
very natural physical explanations in string theory. In particular, we
will focus on two very important features in this subject. The first
one is the natural appearence of a homological, as opposed to
cohomological, framework in which to describe D-brane charge. This point has
been elucidated recently in a variety of different
contexts~\cite{Periwal1}--\cite{AST}. While the standard K-theory
approach does give very natural insights into the physical basis of D-brane
constructions, we shall see that K-homology arises in a far more
physically transparent manner. The second point which we will
emphasize is that the most basic, physical characteristics of D-branes
immediately lead to a definition of their charges in terms of
topological, rather than analytical, K-homology. This description
thereby provides the correct geometric arena for understanding the
physics of D-branes in string theory. For the most part
this will be a review of many well-known results which have arisen
over the last few years from studies of the relationships between
K-theory and D-branes. We will, however, inject some new proposals
concerning certain topological K-homology groups.

The organization of this paper is as follows. In the next section we
will review the standard physical and geometric arguments leading to
the K-theoretic classification of D-brane charges. In section~3 we
will then start showing how these constructs can be very naturally
reinterpreted in terms of analytic K-homology. In section~4 we present
the arguments which imply that D-brane charges are properly classified
in terms of topological K-homology. In section~5 we consider the
effects of curved backgrounds in string theory and the arguments which
imply that an appropriate twisted version of K-theory should classify
D-branes in these instances. Here we also propose a new, physically
motivated definition of twisted topological K-homology groups, which
is also the most natural one from a mathematical perspective.

\newsection{Brane Charges, Superconnections, and
  K-Theory\label{BCSK-T}}

In this section we will review some of the standard relationships between
K-theory and D-branes, emphasizing the geometrical structures
involved. Here we will not give a very precise definition of strings
and D-branes, but instead focus on the mathematical entities which
describe them.

\newsubsection{Brane-Antibrane Systems}

A D-brane may be defined topologically as a relative map
\beq
(\Sigma\,,\,\partial\Sigma)~\stackrel{\chi}{\longrightarrow}~(X\,,\,M) \ ,
\label{Dbranerelmap}\eeq
where $\Sigma$ is a Riemann surface with boundary called a string
worldsheet. We shall work throughout with only ``Type~II
superstrings'', which amounts to assuming that $\Sigma$ is
oriented. The Euclidean spacetime manifold $X$ is ten-dimensional,
oriented and spin, and we assume that
$M\stackrel{\phi}{\hookrightarrow}X$ is a closed, oriented embedded
submanifold. For the time-being we will assume that this embedding is
topologically trivial, and hence that the D-brane worldvolume $M$
carries a spin$^c$ structure~\cite{Witten1,OlsenSz1}. These conditions
will be relaxed in section~5. Among other things, associated with the
D-brane is also a complex vector bundle $\xi\to M$ with connection
$\nabla$, called the Chan-Paton bundle. The rank of this bundle is the
number of coincident D-branes which wrap the common worldvolume $M$.

A brane-antibrane system is defined as such a configuration whereby
the Chan-Paton bundle carries a $\zed_2$-grading
\beq
\xi=\xi^+\ominus\xi^- \ ,
\label{xigraded}\eeq
where the $\pm$ labels indicate the branes and antibranes, all of
which wrap the same worldvolume $M$. The physical requirement
that such a system be invariant under processes involving identical
brane-antibrane creation and annihilation is the mathematical
statement of stable isomorphism of Chan-Paton bundles. Then, physical
quantities which are invariant under deformations of $\xi$ depend only
on the K-theory class $x=[\xi^+]-[\xi^-]\in\K^*(M)$. Since $M$ is
spin$^c$, one can use the Thom isomorphism and the ensuing Atiyah-Bott-Shapiro
construction~\cite{Witten1}--\cite{OlsenSz1} to naturally map $x$
into a class $\K^*(M)\to\K^*(X)$ in
the K-theory of spacetime (or equivalently in the K-theory
$\K^*(N_XM)$ of the normal bundle to $M$ in~$X$). Throughout, when
either $M$ or $X$ is only locally compact, we shall work in K-theory
with compact support, which means physically that we measure D-brane
charge with respect to that of the vacuum.

\begin{figure}[htb]
\epsfxsize=2in
\bigskip
\centerline{\epsffile{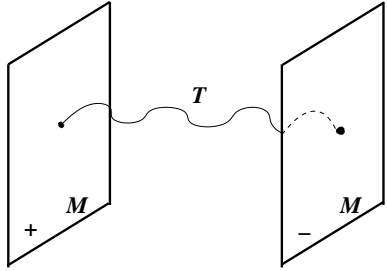}}
\caption{\baselineskip=12pt{\it A
    brane-antibrane system wrapping a common worldvolume $M$. The
    tachyon field $T$ is associated with open string modes which
    stretch between the branes and antibranes.}}
\bigskip
\label{MT+-}\end{figure}

Because of the $\zed_2$-grading, the natural geometric object to
consider on this system of D-branes is not a connection but rather a
superconnection on (\ref{xigraded})~\cite{Witten1,Quillen}
\beq
\A=\pmatrix{\nabla^+&T\cr\overline{T}&\nabla^-\cr} \ ,
\label{superconn}\eeq
where $\nabla^\pm:\xi^\pm\to\xi^\pm$ are connections on the branes and
antibranes, while $T:\xi^+\to\xi^-$ and $\overline{T}:\xi^-\to\xi^+$
are called tachyon fields. The brane-antibrane
system, with its superconnection (\ref{superconn}), is depicted schematically
in fig.~\ref{MT+-}.

\newsubsection{D-Brane Charges}

Determining a formula for D-brane charge is the problem
of finding the chiral gauge and gravitational anomalies induced on the
brane worldvolume $M$ from the restriction of spacetime spinors in
$K_X^{1/2}$~\cite{MinMoore,Szsuper,SchwarzWit}. It amounts to
computing the index of the generalized Dirac operator
\beq
\pmatrix{\ii\SlashN^+&T\cr\overline{T}&\ii\SlashN^-\cr}
\label{genDirac}\eeq
associated to the superconnection (\ref{superconn}), by ``splitting''
the corresponding $\widehat{A}$-class in two. Mathematically, this
procedure consists of introducing a bilinear pairing on rational cohomology
defined by
\beq
\HH^*(X,\rat)\otimes\HH^*(X,\rat)~\longrightarrow~\rat \ , ~~\langle
x,y\rangle=(x\cup y)[X] \ ,
\label{cohpairing}\eeq
and a homomorphism from K-theory to cohomology given by
\beq
G\,:\,\K^*(X)~\longrightarrow~\HH^*(X,\rat) \ , ~~ G(\xi)=\ch(\xi)~
\widehat{A}^{\,1/2}(X) \ ,
\label{Gxidef}\eeq
where
\beq
\ch(\xi)={\rm STr}~\e^{-\A^2/2\pi}\equiv\Tr\pmatrix{\id&0\cr
0&-\id\cr}~\e^{-\A^2/2\pi}
\label{Cherndef}\eeq
is the Chern character of the graded bundle $\xi$. Then the
index theorem gives
\beq
\Bigl\langle G(\xi)\,,\,G(\xi')\Bigr\rangle={\rm
  index}(\xi\otimes\xi') \ .
\label{Gindexsplit}\eeq

Worldvolume integrals of the cohomology class
\beq
Q=\ch(\phi^{~}_!\xi)~\widehat{A}^{\,1/2}(X)
\label{chargecohclass}\eeq
then define the D-brane charge~\cite{Szsuper,SchwarzWit,KenWil},
where $\phi^{~}_!$ is the Gysin map on K-theory induced by the
worldvolume embedding into $X$. Note that when the tachyon field is
turned off, $T=\overline{T}=0$, the Chern character (\ref{Cherndef}) is the
difference $\ch(\xi)=\ch(\xi^+)-\ch(\xi^-)$. The corresponding charge
(\ref{chargecohclass}) is then the sum of the charges of the separated
branes and antibranes.

In the following we shall also have the occasion to deal with systems
of unstable D-branes. They are obtained from brane-antibrane systems
by taking the quotient of the Chan-Paton bundle (\ref{xigraded}) by
the action of the left-moving, worldsheet fermion parity operator
$(-1)^{F_{\rm L}}$, which amounts to eliminating the $\zed_2$-grading
and making the identifications
\beq
\nabla^+=\nabla^- \ , ~~ T=\overline{T}
\label{superconnid}\eeq
in the superconnection (\ref{superconn}). The charge formula is still
given by (\ref{chargecohclass}), but now the supertrace defining the
Chern character is modified to
\beq
{\rm Str}~\e^{-\A^2/2\pi}\equiv{\rm Tr}~\e^{-\A^2/2\pi}\,
\pmatrix{0&\id\cr\id&0\cr} \ .
\label{Chernmod}\eeq

\newsubsection{Tachyon Condensation}

A physical feature of the K-theory classification of D-brane charge is that
stable BPS D-branes may be constructed as solitonic decay products in
the worldvolumes of brane-antibrane
systems~\cite{Witten1}--\cite{OlsenSz1}. For example, let us describe
how to build $k$ D$p$-branes (for odd $p<9$) from $N=2^{(9-p)/2}\cdot k$
  D9--$\overline{\rm D9}$~pairs. The worldvolume $M$ is
  $p+1$-dimensional, and for the Chan-Paton bundle
  (\ref{xigraded}) we take
\beq
\xi^\pm=\Delta^\pm\otimes\xi_k \ ,
\label{CPchoose}\eeq
where $\Delta^\pm$ are the irreducible, chiral spin bundles over the normal
bundle $N_XM$ of rank $2^{(9-p)/2}$, and $\xi_k$ is a trivial complex
vector bundle of rank $k$. It is important to note that, when $M$ is only
spin$^c$, the usual spinor bundles $\sigma^\pm$ do not in general exist
globally and need to be twisted into globally defined vector
bundles over $M$~\cite{Witten1,OlsenSz1,Szsuper}. For this, let
$\ell\to M$ be a complex line bundle whose first Chern class
$c_1(\ell)$ reduces modulo 2 to the second Stiefel-Whitney class
$w_2(N_XM)$. Then the square root $\ell^{1/2}$, with
$\ell^{1/2}\otimes\ell^{1/2}=\ell$, also cannot in general be
constructed globally, because when there is two-torsion in
$\HH^2(M,\zed)$ then there are different square roots of $\ell$, and
hence more than one spin$^c$ structure for a given class
$c_1(\ell)$. However, the twisted spinor bundles
$\Delta^\pm=\ell^{1/2}\otimes\sigma^\pm$ {\it do} exist as vector bundles
over $M$. This is just the precise meaning of the existence of a
spin$^c$ structure, and it implies the vanishing of global worldsheet
anomalies in the case of a topologically trivial
$B$-field background~\cite{FreedWitten1}. We will return to these
issues in section~5.

We choose a coordinate chart on $X$ in which the connections on the
branes and antibranes decompose as
\beq
\nabla^\pm=\sum_{m=1}^{p+1}\id_{\Delta^\pm}\otimes\nabla_m~\dd
x^m+\sum_{i=p+2}^{10}\id_{\Delta^\pm}\otimes
\frac\partial{\partial x^i}~\dd x^i \ ,
\label{nablapmdecomp}\eeq
where $m=1,\dots,p+1$ labels directions along the worldvolume $M$ and
$i=p+2,\dots,10$ labels coordinates in the normal bundle to $M$ in
$X$. For the tachyon profile we use the local representation
\beq
T_t=\frac1{\sqrt t}\,\sum_{i=p+2}^{10}\Bigl[x^i\,\gamma^{~}_i\otimes
\id_{k\times k}-\gamma^{~}_i\otimes\Phi^i(x^m)\Bigr] \ ,
\label{Tprofile}\eeq
where $t$ is a positive parameter labelling a family of tachyon
fields. The first term in (\ref{Tprofile}) situates the D-branes at the
spacetime location $x^i=0$, where $\gamma^{~}_i$ generate the
$9-p$-dimensional Clifford algebra in the Majorana-Weyl
representation. It smears the brane charge over a region of size
$\sqrt t$ and provides an off-shell interpolation in string
theory. The fields $\Phi\in\Gamma(M,N_XM\otimes{\rm End}(\xi_k))$ extend
the maps $M\stackrel{\phi}{\hookrightarrow}X$ to sections of the
normal bundle twisted by the endomorphism bundle ${\rm
  End}(\xi_k)$. They represent the fluctuations of the relative
positions of the $k$ D$p$-branes.

The Chern character (\ref{Cherndef}) corresponding to
(\ref{nablapmdecomp}) and (\ref{Tprofile}) is independent of
$t\in\real_+$~\cite{Quillen}. Taking the ``on-shell'' limit $t\to0$ yields the
symmetrized trace formula~\cite{Szsuper,AGS}
\bea
\ch(\xi)&=&\Tr\,{\rm Sym}~\exp-\frac1{2\pi}\,\left(\,\sum_{m,n=1}^{p+1}
\left[\nabla_n\,,\,\nabla_m\right]~\dd x^n\wedge\dd x^m\right.\nn\\&&\left.+
\,\sum_{m=1}^{p+1}\left[\nabla_m~\dd x^m\,,\,\imath^{~}_\Phi\right]+
\left[\imath^{~}_\Phi\,,\,\imath^{~}_\Phi\right]\right) \ ,
\label{chonshell}\eeq
where $\imath^{~}_\Phi$ is internal multiplication with $\Phi$
regarded as a vector field in the transverse space. When substituted
into the formula for D-brane charge, the $\imath^{~}_\Phi$ terms in
(\ref{chonshell}) yield the Myers terms~\cite{Myers} which lead to the
``dielectric effect''. The simplest example of this process is a
D0-brane blowing up into a spherical D2-brane, or in other words into
a D0--D2~bound state.

\newsection{Tachyon Fields and Analytic K-Homology}

In this section we shall analyse the topology of the tachyon fields
introduced in the previous section in some more detail, and show how
they alone lead immediately to the relationship between K-theory and
D-branes. After presenting some arguments supporting a homological
classification of D-brane charge, we will then show that these same
ingredients very naturally give an alternative description in terms of
analytic K-homology.

\newsubsection{Tachyons are Fredholm Operators}

Let $\xi\to M$ be a complex Chan-Paton vector bundle over an unstable
D-brane worldvolume, with structure group $U(N)$. The tachyon field
$T$ is adjoint-valued, so it is a map $T:X\to u(N)$ into the Lie
algebra of $U(N)$. The Lie algebra is a contractible space, so $T$ carries no
topology. A map on $X\to u(N)$ cannot represent an element of
$\K^*(X)$, because it does not carry any topological information at
all. The solution to this problem~\cite{Hori,Witten2} is to set $N=\infty$ and
use infinitely many unstable D-branes. We then interpret the structure
group $U(\infty)$ as the group $U({\cal H})$ of unitary operators on
the separable Chan-Paton Hilbert space ${\cal H}=L^2(M,\xi)$ on which the
open string zero modes act. Whether or not this makes sense
physically, it can always be done mathematically. From a physical
standpoint we may assume that we start with an infinite set of
unstable D-branes and reduce it by tachyon condensation to a finite
set representing a configuration of finite total energy.

Requiring that the system have finite energy and that the induced
D-brane charge be finite is tantamount to assuming that 0 is not an
accumulation point of the spectrum of the self-adjoint operator
$T$. In other words, $\tilde T=T/\sqrt{\id+T^2}\in{\cal B}({\cal H})$ is
a bounded linear operator on $\cal H$ such that $\tilde
T^2-\id\in{\cal K}({\cal H})$~\cite{WeggeOlsen}, where ${\cal K}({\cal
  H})$ is the elementary $C^*$-algebra of compact operators on $\cal H$. These
conditions simply mean that the tachyon field is now a map
\beq
T\,:\,X~\longrightarrow~{\cal F}({\cal H})
\label{TXFredholm}\eeq
into the space ${\cal F}({\cal H})$ of Fredholm operators on the
Chan-Paton Hilbert space.

Carrying up this argument to brane-antibrane systems, the homotopy
classes of maps (\ref{TXFredholm}) yield an epimorphism
\beq
\Bigl[X\,,\,{\cal F}({\cal H})\Bigr]~\longrightarrow~\K^*(X)~\longrightarrow~0
\label{XcalFKX0}\eeq
which is given by taking the index bundle whose fiber over a point
$x\in X$ is\footnote{\baselineskip=12pt More precisely, there is an
  isomorphism $[X,{\cal F}({\cal H})]\to\K^0(X)$. The analogous
  result for unstable D-branes yields an isomorphism $[X,{\cal F}^{\rm
    skew}({\cal H})]\to\K^{-1}(X)$, where ${\cal F}^{\rm skew}({\cal
    H})$ is the space of skew-adjoint Fredholm operators on $\cal H$
  (or, more precisely, its component consisting of operators which
  are not essentially positive or negative). The analog of
  (\ref{Indexbundle}) for this latter map may be found in~\cite{Hori}.}
\beq
{\rm Index}(T)_x=\ker T(x)\ominus{\rm coker}\,T(x) \ .
\label{Indexbundle}\eeq
Then $[{\rm Index}(T)]$ is the K-theory class of the D-brane
Chan-Paton space. Note that in this case the Fredholm operator $T$ can
without loss of generality be taken to be a partial
isometry~\cite{HarveyMoore1}, i.e. $T\,\overline{T}\,T=T$, which is
the basis of the representation of D-branes as noncommutative solitons
in string field theory~\cite{HKLM1}.

\newsubsection{Tachyons are Classifying Maps}

A somewhat more direct argument comes from minimizing the tachyon
potential $V(T)$. The equations
\beq
\dd T=0 \ , ~~ V'(T)=0
\label{mineqns}\eeq
are solved by
\beq
T=\sum_n\lambda_n\,P_n \ ,
\label{Tsolve}\eeq
where $P_n$ are orthogonal projection operators on $\cal H$ and
$V'(\lambda_n)=0$. The basic shape of the tachyon potential is such
that there are two stationary points~\cite{HKLM1}. A global minimum
$V(0)=0$ occurs at $\lambda=0$ representing the closed string vacuum, and an
extremum appears at $\lambda=t_*$ with $V(t_*)$ giving the tension of the
D-brane in the perturbative open string vacuum. Thus the
only non-trivial solution is $T=t_*\,P_n$, where $P_n$ is a projection
operator of rank $n$.

It follows that slowly-varying tachyonic field configurations on $X$
are given by maps
\beq
T\,:\,X~\longrightarrow~BU(n)
\label{TXBUn}\eeq
into the space $BU(n)$ of rank $n$ projectors on the Chan-Paton
Hilbert space. But $BU(n)$ is also the classifying space for complex
vector bundles of rank $n$, so that
\beq
{\rm Vect}_n(X)\cong\Bigl[X\,,\,BU(n)\Bigr] \ .
\label{BUnclassifying}\eeq
Thus homotopy classes of tachyons are directly related to K-theory.

\newsubsection{D-Brane Charge Lives in K-Homology}

When the charges of D-branes are defined by the behaviour of
Ramond-Ramond fields, which are differential forms on $X$, K-theory is
the appropriate contravariant cohomology theory to
use~\cite{MooreWitten1}. A continuous map $\varphi:X\to X'$ induces a
pull-back on K-theory, $\varphi^*:\K^*(X')\to\K^*(X)$. However,
D-branes also represent supergravity solutions which are cycles in $X$
corresponding to the worldvolumes of D-branes. They should therefore be
properly classified by a {\it homology}
theory~\cite{Periwal1}. K-homology carries information of both cycles
in $X$ and of gauge bundles over them, and it is naturally implied by
Poincar\'e duality for spin$^c$ manifolds $M$. As we shall
discuss, D-brane charge is most naturally regarded as an element of
K-homology, rather than K-theory, because:
\begin{itemize}
\item D-branes {\it a priori} carry stable vector bundles, rather than
  virtual bundles.
\item K-homology transforms charges covariantly under maps
  $\varphi:X\to X'$ of the spacetime, i.e. if $\phi:M\to X$ is a
  D-brane, then its push-forward $M\mapsto\varphi^{~}_*(M)\subset X'$
  gives rise to a map $\varphi^{~}_*:\K_*(X)\to\K_*(X')$ on
  K-homology. This makes contact with the covariant, operational
  definition of D-branes as submanifolds $M$ of the spacetime~$X$.
\item D-brane charge in the brane-antibrane constructions is
  determined in a very precise way from the K-theory group of the
  worldvolume normal bundle.
\end{itemize}

K-homology is the natural dual classification tool to K-theory. The
index map provides a natural bilinear pairing
\beq
\K^*(X)\times\K_*(X)~\stackrel{\rm index}{\longrightarrow}~\zed \ .
\label{Khompairing}\eeq
Furthermore, the K-theory lift of the Poincar\'e duality
relation $\HH_*(X,\zed)\cong\HH^{10-*}(X,\zed)$ yields
\beq
\K_*(X)\cong\K^{10-*}(X) \ .
\label{Kduality}\eeq
The relationship (\ref{Kduality}) has a very natural physical
interpretation. Its right-hand side represents the construction of
D-branes from non-BPS D9-brane configurations, which classifies
D-brane charge defined by Ramond-Ramond fields on $X$. The left-hand
side of (\ref{Kduality}) represents the construction of D-branes from
non-BPS D-instantons~\cite{ter1} and classifies the D-brane
worldvolume embedded into the spacetime manifold $X$.

\newsubsection{Tachyonic Configurations Live in Analytic K-Homology}

Motivated by the arguments of the previous subsection, let us now turn
the analysis of the beginning of this section around somewhat to show
that tachyon fields naturally lead to the relationship between
D-branes and K-homology. For simplicity and ease of notation, we will
assume that the D-brane worldvolume $M$ is of odd dimension and focus
on the tachyon field $T=\overline{T}$ of a system of unstable branes. The
corresponding result for brane-antibrane systems and $M$ of even
dimension is easily obtained by grading all of the structures which
follow. The basic observation is that the tachyon field $T$ (or more precisely
its bounded extension $\tilde T$), along with the worldvolume
embedding extension $\phi\mapsto\Phi$ on the Chan-Paton Hilbert space
$\cal H$, can be assembled into a ``Fredholm module'' $({\cal H},\Phi,T)$
which is specified by the following pieces of data~\cite{WeggeOlsen}:
\begin{itemize}
\item $\cal H$ is the separable Chan-Paton Hilbert space.
\item $\Phi:C(M)\to{\cal B}({\cal H})$ is a $*$-homomorphism. Here we
  have used Gel'fand duality to replace the manifold $M$ by the
  $C^*$-algebra $C(M)$ of continuous complex-valued functions on $M$. This
  map is responsible for the representation of D-branes in terms of
  non-BPS D-instantons.
\item $T\in{\cal B}({\cal H})$ is a self-adjoint operator obeying the
  finite energy conditions $T^2-\id\in{\cal K}({\cal H})$ and
  $[T,\Phi]\in{\cal K}({\cal H})$.
\end{itemize}

Then the analytic K-homology $\K_*(M)=\K^*(C(M))$ is the group of
equivalence classes of Fredholm modules with respect to the
equivalence relation $\sim$ generated by the following three
relations:
\begin{itemize}
\item $\underline{\it Gauge~Symmetry:}$ For any unitary operator
  $U\in{\cal B}({\cal H},{\cal H}')$,
\beq
({\cal H}\,,\,\Phi\,,\,T)\sim({\cal H}'\,,\,U\,\Phi\,U^{-1}\,,\,
U\,T\,U^{-1}) \ .
\label{gaugeequiv}\eeq
\item $\underline{\it
    Annihilation~of~Virtual~Non-BPS~D-Instantons~by~Tachyon~Condensation:}$\\
    Define a degenerate Fredholm module $({\cal H}',\Phi',T')_{\rm
    deg}$ as one with a trivial tachyonic field configuration,
    i.e. $T'^2-\id=[T',\Phi']=0$. Then for any degenerate Fredholm
    module,
\beq
({\cal H}\,,\,\Phi\,,\,T)\oplus({\cal H}'\,,\,\Phi'\,,\,T')_{\rm
    deg}\sim({\cal H}\,,\,\Phi\,,\,T)
\label{degadd}\eeq
for all Fredholm modules $({\cal H},\Phi,T)$.
\item $\underline{\it Continuous~Deformations~of~Tachyons:}$ Any two
  Fredholm modules $({\cal H},\Phi,T)$ and $({\cal
  H}',\Phi',T')$ are equivalent if there is an operator homotopy
  ${\cal H}\sim{\cal H}'$, $\Phi\sim\Phi'$, and there exists a norm
  continuous path connecting $T$ and $T'$. This condition may be
  regarded as the statement of D-brane charge conservation in physical
  processes.
\end{itemize}
We further require that each equivalence class be unaffected by
continuous deformations of $\cal H$ and $\Phi$.

\newsection{D-Branes and Topological K-Homology}

In the previous section we arrived at the conclusion that D-brane
charge, generated through tachyon condensation on non-BPS brane
systems, is most naturally described within the framework of Fredholm
modules and analytic K-homology. However, this approach, like other
functional analytic approaches to
K-homology~\cite{Periwal1}--\cite{AST}, obscure the inherent geometry
of D-branes. In this section we will accomplish two things. First, we
will show that D-branes can be described directly in the language of
Fredholm modules, without the need of resorting to their description
in terms of tachyon fields on non-BPS configurations. Second, this
formulation will immediately lead us into a more geometrical framework
for analysing D-brane charges and their properties, which will turn
out to be the natural physical arena.

\newsubsection{D-Branes are Fredholm Modules}

We will begin by demonstrating that every D-brane provides a Fredholm
module over the algebra $C(X)$ of functions on spacetime. Let
$\phi:M\to X$ be a closed, odd dimensional spin$^c$ submanifold of the
spacetime $X$, representing the brane worldvolume (Again the result for even
dimensional $M$ is easily obtained by introducing appropriate
$\zed_2$-gradings). $M$ is equipped with a complex Chan-Paton vector
bundle $\xi\to M$ with connection $\nabla$, and it inherits a metric from
$X$. From the spinor representation of the spin$^c$ group, we can
define a (twisted) spinor bundle $\Delta\to M$ and the associated
separable Hilbert space
\beq
{\cal H}=L^2(M,\Delta\otimes\xi)
\label{calHMtildeS}\eeq
of square-integrable spinors on $M$ with values in $\xi$. Let
$\Dirac_\xi$ be the usual twisted Dirac operator on
$\Gamma(M,\Delta\otimes\xi)$ with respect to the chosen connection on
the bundle $\Delta\otimes\xi$. Then $\Dirac_\xi$ is an unbounded self-adjoint
operator on the Hilbert space (\ref{calHMtildeS}). Let us now define a
representation $\Phi:C(X)\to{\cal B}({\cal H})$ as pointwise
multiplication on (\ref{calHMtildeS}) by the function
\beq
\Phi(f)=f\circ\phi \ , ~~ \forall f\in C(X) \ .
\label{PhirepCX}\eeq
This construction produces an (unbounded) Fredholm module $({\cal
  H},\Phi,\Dirac_\xi)$.

The Fredholm module just built defines an element of K-homology $\K_*(X)$
  which is independent of the choice of connection on
  $\Delta\otimes\xi$. Moreover, it can be shown that all classes in
  $\K_*(X)$ can be obtained from this construction by using an
  appropriate D-brane $(M,\phi,\xi)$~\cite{BaumDoug}. By putting a
  suitable equivalence relation on the set of all such triples
  $(M,\phi,\xi)$, the classes in $\K_*(X)$ are in a bijective
  correspondence with the corresponding equivalence classes
  $[(M,\phi,\xi)]$. It is precisely this correspondence that now leads
  us into the geometrical formulation of D-brane charge.

\newsubsection{D-Brane Charge Lives in Topological K-Homology}

The fact, discussed in the previous section, that the natural setting
in which D-brane charges emerge should be topological motivates the
development of a geometric approach to K-homology. For this, consider the free
abelian group generated by certain triples
\beq
k(X)={\rm Span}^{~}_\zeds\left\{(M\,,\,\phi\,,\,x)~\left|~
{\begin{array}{l}
M~{\rm closed,}~{\rm spin}^c\\\phi\,:\,M~\longrightarrow~X\\
x\in\K^*(M)\end{array}}\right.\right\} \ ,
\label{kXtriples}\eeq
where in each triple $(M,\phi,x)$ it is understood that the manifold $M$ is
equipped with a {\it fixed} spin$^c$ structure on its tangent bundle $TM$.
Each triple in $k(X)$ will be refered to as a homology K-cycle,
because they each naturally give rise to a K-homology class in
$X$. The reason for this is very simple. Since $M$ is spin$^c$, we can
``cheat'' by exploiting the fact that $M$ satisfies Poincar\'e duality
in K-theory. Thus the class $x\in\K^*(M)$ has a Poincar\'e dual in the
K-homology $\K_*(M)$ of $M$, which under the push-forward
$\phi_*:\K_*(M)\to\K_*(X)$ can be mapped to a class in the K-homology
of spacetime $X$. Thus there is a natural homomorphism $k(X)\to\K_*(X)$.

This is the geometric description of K-homology, which we shall refer
to as topological K-homology to distinguish it from its functional
analytic counterpart. It is clear then that any collection of D-branes
with worldvolume $\phi:M\to X$ and Chan-Paton bundle $\xi$ naturally
defines an element of the topological K-homology
$(M,\phi,[\xi])\in\K_*(X)$ of spacetime. The two main, natural physical
properties of this description are~\cite{BaumDoug}:
\begin{itemize}
\item $\underline{\it
    Gauge~Symmetry~Enhancement~for~Coincident~D-Branes:}$ The
    disjoint union of two coincident D-brane triples gives
\beq
\Bigl(M\,,\,\phi\,,\,[\xi]\Bigr)\amalg\Bigl(M\,,\,\phi\,,\,[\xi']\Bigr)
\cong\Bigl(M\,,\,\phi\,,\,[\xi\oplus\xi']\Bigr) \ .
\label{symenhance}\eeq
\item $\underline{\it D-Brane~Charges:}$ The Chern character in
  K-homology may be defined as the rational homology class
\beq
\ch(M,\phi,x)=\phi_*\Bigl(\ch(x)\cup\widehat{A}(M)\cap[M]\Bigr) \ .
\label{chhomology}\eeq
\end{itemize}

\begin{figure}[htb]
\epsfxsize=4in
\bigskip
\centerline{\epsffile{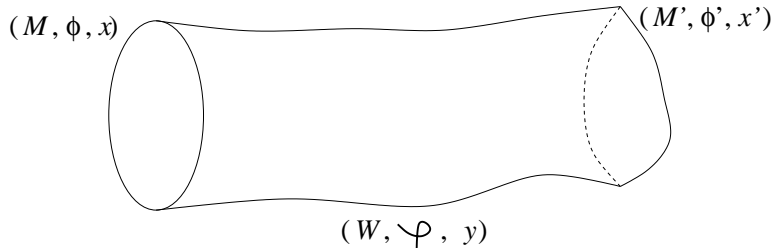}}
\caption{\baselineskip=12pt{\it D-brane charge conservation in
    topological K-homology is represented through bordism.}}
\bigskip
\label{bordism}\end{figure}

The map $k(X)\to\K_*(X)$ is surjective, and its kernel is generated by two
relations called ``bordism'' and ``vector bundle modification''. These
two equivalence relations required to quotient $k(X)$ have very
natural interpretations in {\it stable} D-brane
physics~\cite{HarveyMoore1,AST}. The bordism relation is
straightforward to describe. It is depicted in fig.~\ref{bordism} and
is essentially the requirement of D-brane charge conservation. Namely,
two triples are equivalent,
\beq
(M\,,\,\phi\,,\,x)\sim(M'\,,\,\phi'\,,\,x') \ ,
\label{bordismtriple}\eeq
if there is a K-cycle $(W,\varphi,y)$ which interpolates between the
triples in the sense that
\beq
\left(\partial W\,,\,\varphi\Bigm|_{\partial W}\,,\,y\Bigm|_{\partial
    W}\right)\cong(M\,,\phi\,,\,x)\amalg(-M'\,,\,\phi'\,,\,x') \ .
\label{Wbordism}\eeq
Here $W$ is a spin$^c$ manifold with boundary, $y=[\xi]$ with
$\xi\to W$ a complex vector bundle, $\varphi:W\to X$ is continuous,
and $-M'$ denotes the submanifold $M'$ with the reversed spin$^c$
structure. This relation represents ``continuous'' deformations of the
D-brane worldvolume together with the Chan-Paton gauge bundle over
it. The other relation is somewhat more involved and is described in
the next subsection. It describes a well-known descent relation in D-brane
physics which is due to the non-abelian nature of gauge bundles over
multiple D-brane configurations.

\newsubsection{Vector Bundle Modification is the Dielectric Effect}

Vector bundle modification is the relation that identifies the triples
\beq
(M\,,\,\phi\,,\,x)\sim\Bigl(\hat M\,,\,\phi\circ\rho\,,\,[\hat\xi]
\otimes\rho^*(x)\Bigr) \ ,
\label{vectormod}\eeq
where $\hat M\stackrel{\rho}{\to}M$ is a sphere bundle over $M$ whose
fiber $\sphere_p=\rho^{-1}(p)$, $p\in M$ is a sphere of even dimension
$2n$, and $\hat\xi\to\hat M$ is a vector bundle over $\hat M$ such
that for all $p\in M$, $[\hat\xi]|_{\spheres_p}$ is the generator of the
reduced K-theory group $\tilde\K^*(\sphere_p)=\zed$. As we will now
demonstrate, the relation (\ref{vectormod}) identifies a spherical
D-brane, carrying a non-trivial gauge bundle, with a lower dimensional
D-brane. It therefore represents the K-homology manifestation of the
dielectric effect~\cite{HarveyMoore1,AST,Myers} that we mentioned in section~2.

Let $\xi\to M$ be a spin$^c$ vector bundle of rank $2n$, and define
\beq
\hat M=\ball(\xi)_+\,{\bigcup}_{\spheres(\xi)}\,\ball(\xi)_- \ ,
\label{hatMball}\eeq
where $\ball(\xi)_\pm$ are two copies of the unit ball bundle of $\xi$
whose boundary $\partial\ball(\xi)=\sphere(\xi)$ is the unit sphere bundle of
$\xi$. The copies are glued together using the identity map on
$\sphere(\xi)$, so that (\ref{hatMball}) is a sphere bundle over the
original D-brane worldvolume $M$, with fiber of dimension $2n$, which
is also a spin$^c$ manifold. The bundles $\ball(\xi)_\pm$ are interpreted
as the worldvolumes of D$(2n+m)$-branes and $\overline{{\rm
    D}(2n+m)}$-branes, where $m=\dim(M)-1$. When glued together, $\hat
M$ becomes the worldvolume of a spherical D$(2n+m)$-brane wrapped on a
$2n$-dimensional sphere.

Since $\xi$ has even rank, we can introduce the pull-backs
  $\Delta^\pm(\xi)$ of the associated chiral spinor bundles to $\xi$ using
  the bundle projection $\xi\to M$. We interpret $\Delta^\pm(\xi)$ as the
  Chan-Paton vector bundles on the D$(2n+m)$-branes and $\overline{{\rm
  D}(2n+m)}$-branes. The graded Chan-Paton bundle on this
  brane-antibrane system is therefore given by
\beq
\hat\xi=\Delta^+(\xi)\Bigm|_{\balls(\xi)_+}\,
{\bigcup}_{T\bigm|_{\spheress(\xi)}}\,\Delta^-(\xi)
\Bigm|_{\balls(\xi)_-} \ ,
\label{hatxidef}\eeq
where the gluing is done using the transition function $T$ on
$\sphere(\xi)$ defined by
\beq
T(p,v)=\sum_{l=1}^{2n}v^l\,\gamma^{~}_l
\label{transfnsphere}\eeq
with $v$ a unit vector in the $2n$-dimensional vector space
which is the fiber of $\sphere(\xi)$ at $p\in M$, and $\gamma^{~}_l$ are
positive chirality $SO(2n)$ gamma-matrices. We interpret
(\ref{transfnsphere}) as the tachyon field created by open strings
stretched between the D$(2n+m)$-branes and the $\overline{{\rm
    D}(2n+m)}$-branes. After tachyon condensation, the tachyonic
configuration (\ref{transfnsphere}) thereby induces D$m$-brane
charge. It follows that the configuration $(\hat
M,\phi\circ\rho,[\hat\xi]\otimes\rho^*(x))$ should then be physically
identified with the D$m$-branes characterized by the K-cycle
$(M,\phi,x)$.

\newsection{$B$-Fields and Twisted K-Homology\label{BFieldTwist}}

All of our analysis thus far has been implicitly done in flat
space. We now consider the generalizations of the previous results to
curved string backgrounds. By the string equations of motion, this is
tantamount to making a certain supergravity field, called the
``$B$-field'', non-trivial. We will begin with a purely mathematical
explanation of what a $B$-field is, and then show that its presence
implies that D-brane charge should take values in a certain twisted
version of K-theory. The development of twisted K-theory is currently
at the very heart of recent activity in this
field~\cite{Witten1,HarveyMoore1},\cite{Kapustin}--\cite{Freed1}. The
main result of this section will be a new proposal~\cite{STWunpub},
which we present here without proof, for twisted topological
K-homology groups. With this proposal the physical characteristics of
twisted K-theory groups become transparent, and it provides the basis
for the geometrical characterization of D-branes in curved backgrounds.

\newsubsection{$B$-Fields}

A $B$-field may be defined geometrically as a gerbe with
1-connection~\cite{Kapustin}. Let us first define precisely what we
mean by a gerbe~\cite{Bryl} in a form that will be most useful in the
following. Recall that integer cohomology groups form a hierarchy that
may be used to topologically classify geometrical structures over the
spacetime manifold $X$. For instance, $\HH^1(X,\zed)$ classifies
functions on $X\to\sphere^1$, while $\HH^2(X,\zed)$ provides
characteristic classes for complex line bundles over $X$. Gerbes yield
geometric realizations for the degree three cohomology classes of
$\HH^3(X,\zed)$.

For any group $G$, let $\underline{G}_{\,X}$ denote the corresponding
constant sheaf over $X$ of locally constant $G$-valued functions. Then
the sheaf cohomology $\HH^1(\underline{G}_{\,X})\cong\HH^1(X,G)\cong[X,BG]$
classifies principal $G$-bundles over $X$. Let us look at some
examples of this characterization which will be particularly important
for the analysis which follows. Consider the exact sequence
\beq
0~\longrightarrow~\zed~\longrightarrow~\complex~\stackrel{\exp}
{\longrightarrow}~\complex^\times~\longrightarrow~0 \ .
\label{complexsheaf}\eeq
Since $\underline{\complex}_{\,X}$ is a soft sheaf over $X$, the
corresponding long exact sequence in sheaf cohomology yields an
isomorphism
\beq
\HH^*(\underline{\complex^\times}_{\,X})~\stackrel{\sim}{\longrightarrow}~
\HH^{*+1}(\underline{\zed}_{\,X})~\cong~\HH^{*+1}(X,\zed) \ .
\label{longexactcomplex}\eeq
In particular,
$\HH^2(X,\zed)\cong\HH^1(\underline{\complex^\times}_{\,X})\cong[X,BU(1)]$,
as expected.

Next, if $\cal H$ is a separable Hilbert space,
then there is an exact sequence
\beq
0~\longrightarrow~U(1)~\longrightarrow~U({\cal H})~
\longrightarrow~PU({\cal H})~\longrightarrow~0 \ .
\label{Hilbertsheaf}\eeq
By Kuiper's theorem~\cite{Kuiper}, the unitary group $U({\cal H})$ is
contractible, and so it yields soft sheaves with no cohomology. Thus again we
generate an isomorphism
\beq
\HH^1\Bigl(\underline{PU({\cal H})}_{\,X}\Bigr)~\stackrel{\sim}
{\longrightarrow}~\HH^2(\underline{\complex^\times}_{\,X})~\cong~
\HH^3(X,\zed)
\label{longexactHilbert}\eeq
where we have in addition used (\ref{longexactcomplex}). It follows
that the third integer cohomology group
\beq
\HH^3(X,\zed)\cong\Bigl[X\,,\,BPU({\cal H})\Bigr]
\label{H3class}\eeq
classifies principal $PU({\cal H})$-bundles over the spacetime $X$. For
a given such bundle, the corresponding isomorphism class in
$\HH^3(X,\zed)$ is known as the Dixmier-Douady class and it represents
the gerbe theoretic analog of the first Chern class for line bundles.

As our final example, consider
\beq
0~\longrightarrow~U(1)~\longrightarrow~Spin^c(n)~\longrightarrow~
SO(n)~\longrightarrow~0 \ .
\label{spinsheaf}\eeq
Then the obstruction to lifting any principal $SO(n)$-bundle
$SO(n)\hookrightarrow\xi\to X$ to a spin$^c$-bundle over $X$ is a gerbe
\beq
[\xi]\in\HH^1\Bigl(\underline{SO(n)}_{\,X}\Bigr)~\longmapsto~
W_3(\xi)\in\HH^2(\underline{\complex^\times}_{\,X})\cong\HH^3(X,\zed) \ .
\label{longexactspin}\eeq
The obstruction to a spin$^c$ structure on the manifold $X$ is the
third Stiefel-Whitney class $W_3(\xi)\in\HH^3(X,\zed)$, which coincides
with the Dixmier-Douady invariant of the bundle-lifting gerbe.

\begin{figure}[htb]
\epsfxsize=2in
\bigskip
\centerline{\epsffile{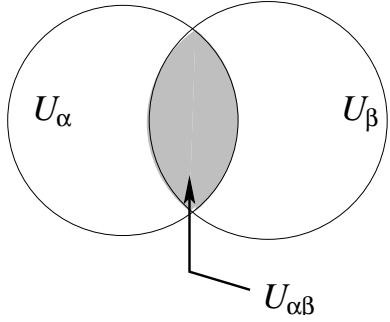}}
\caption{\baselineskip=12pt{\it A good covering of spacetime $X$ by open
    sets $U_\alpha$.}}
\bigskip
\label{covering}\end{figure}

Let us now introduce the notion of connections on gerbes. For this, we
will use a local formulation in terms of bundle gerbes, which provide
a concrete description of a gerbe
$\HH^2(\underline{\complex^\times}_{\,X})$ in the language of
\v{C}ech-de~Rham cohomology. Consider a good cover of the manifold
$X$ by open sets $U_\alpha$ with overlaps $U_{\alpha\beta}=U_\alpha\cap
U_\beta$ (fig.~\ref{covering}). Instead of defining transition
functions on each overlap $U_{\alpha\beta}$, we introduce complex line
bundles $\ell_{\alpha\beta}\to U_{\alpha\beta}$ which satisfy the conditions
\beq
\ell_{\alpha\beta}\otimes\ell_{\beta\gamma}=\ell_{\alpha\gamma}
\label{Lconditions}\eeq
on triple overlaps $U_\alpha\cap U_\beta\cap U_\gamma$. We define a
0-connection to be a connection $\nabla_{\alpha\beta}$ on each line
bundle $\ell_{\alpha\beta}\to U_{\alpha\beta}$. A 1-connection is then defined
to be a 2-form $B_\alpha$ on each $U_\alpha$ such that
\beq
B_\alpha-B_\beta=(\nabla_{\alpha\beta})^2
\label{BLambda}\eeq
on $U_{\alpha\beta}$. The connection must satisfy the local $U(1)$ cocyle
conditions
\beq
\nabla_{\alpha\beta}+\nabla_{\beta\gamma}+\nabla_{\gamma\alpha}=
-\ii~\dd\ln(\zeta_{\alpha\beta\gamma})
\label{Lambdacocycle}\eeq
on triple overlaps, where $\zeta_{\alpha\beta\gamma}$ are
$U(1)$-valued 0-forms obeying
\beq
\zeta_{\alpha\beta\gamma}\,\zeta_{\beta\gamma\lambda}^{-1}\,
\zeta_{\alpha\gamma\lambda}\,\zeta_{\alpha\beta\lambda}^{-1}=1
\label{zetacocycle}\eeq
on quadruple overlaps. The de~Rham representative $H=\dd B$ of the
curvature of the $B$-field then gives rise to an integer cohomology class
$[H]\in\HH^3(X,\zed)$ which coincides with the Dixmier-Douady class of
the bundle gerbe (Of course $[H]$ vanishes in de~Rham cohomology
$H_{\rm DR}^3(X)$). Combining this with the description above in terms
of $PU({\cal H})$-bundles, we may alternatively view the $B$-field as
a connection $\nabla:{\cal A}\to{\cal A}$ on the algebra ${\cal
  A}=C(X)\otimes{\cal K}({\cal H})$, with ${\rm Aut}({\cal K}({\cal
  H}))=PU({\cal H})$ the group of projective unitary automorphisms of
the Hilbert space~$\cal H$. The 0-connections in this setting
are derivations on the endomorphism bundle of the given principal
$PU({\cal H})$-bundle.

\newsubsection{D-Branes in $B$-Fields}

To understand the effect of a non-trivial $B$-field background on the
geometry of D-branes, we will describe the Freed-Witten
anomaly~\cite{FreedWitten1} which arises due to a sign ambiguity in
the open string worldsheet functional integral. The result of
integrating over worldsheet fermions in
$\Gamma(\Sigma,K_\Sigma^{1/2}\otimes\chi^*(TX))$ produces a determinant
of the corresponding Dirac operator $\Dirac$ which defines a section
of the Pfaffian line bundle Pfaff. The pertinent terms are given in
the product
\beq
{\rm Pfaff}(\Dirac)~\cdot~{\rm Hol}_\Sigma(B)~\cdot~
{\rm Hol}_{\partial\Sigma}(\nabla) \ .
\label{Pfaffprods}\eeq
When $\partial\Sigma\neq\emptyset$, the gerbe holonomy is not
invariant under gauge transformations (\ref{BLambda}) and must be
regarded as a section of a complex line bundle ${\cal L}_B$ with
connection over the free loop space $LM$ of the D-brane worldvolume
$M$. The holonomy of the Chan-Paton gauge connection $\nabla$ produces
a trivialization of the twisted Pfaffian line bundle ${\rm
  Pfaff}\otimes{\cal L}_B$ over $LM$. In order for the functional
integral to be globally well-defined, we must therefore require that
the bundle ${\rm Pfaff}\otimes{\cal L}_B|_{LM}$ be topologically trivial.

To understand what this means, let us consider a family of open string
worldsheets $\Sigma_t$, $0\leq t\leq1$ ending on $M$, with
$\Sigma_0=\Sigma_1=\Sigma$. Then, as $t$ varies, the family
$\partial\Sigma_t$ sweeps out a 2-cycle $C_2\subset M$. The holonomy
of the fermion determinants around a circle $\sphere^1$ is the sign
factor~\cite{FreedWitten1}
\beq
\exp\left(\ii\pi\,\Bigl\langle w_2(M)\,,\,[C_2]\Bigr\rangle\right) \ ,
\label{holonomysign}\eeq
where $w_2(M)\in\HH^2(M,\zed_2)$ is the second Stiefel-Whitney class
which may be related to the third Stiefel-Whitney class introduced in
the previous subsection as follows. Consider the exact sequence
\beq
0~\longrightarrow~\zed~\stackrel{\times2}{\longrightarrow}~\zed~
\stackrel{r}{\longrightarrow}~\zed_2~\longrightarrow~0 \ ,
\label{zedexact}\eeq
where the map $r$ is reduction modulo 2. This induces a long exact
sequence in cohomology
\beq
\cdots~\longrightarrow~\HH^2(M,\zed)~\stackrel{r^*}{\longrightarrow}~
\HH^2(M,\zed_2)~\stackrel{\beta}{\longrightarrow}~\HH^3(M,\zed)~
\longrightarrow~\cdots \ ,
\label{Bocksteinexact}\eeq
where the map $\beta$ is called the Bockstein homomorphism. Then the
image of the second Stiefel-Whitney class under the Bockstein map
yields the desired class, $W_3(M)=\beta(w_2(M))\in\HH^3(M,\zed)$. Note
that since the spacetime $X$ is oriented and spin, we further have
$W_3(M)=W_3(N_XM)$~\cite{Witten1}.

A flat $B$-field has torsion characteristic class $[H]$ which may be
computed from the Bockstein map
$\beta:\HH^2(X,\real/\zed)\to\HH^3(X,\zed)$~\cite{FreedWitten1}. The
holonomy of this class around an $\sphere^1$ yields the first Chern
class of the line bundle ${\cal L}_B$ on the loop space $LX$. We
therefore have
\beq
\oint\limits_{\partial\Sigma}\phi^*\Bigl([H]\Bigr)=c_1({\cal L}_B)
\label{c1calLB}\eeq
over $LM$. On the other hand, the Chern class of the Pfaffian line
bundle over $LM$ may be computed as
\beq
c_1({\rm Pfaff})=\beta\left(~\oint\limits_{\partial\Sigma}w_2(M)
\right)=\oint\limits_{\partial\Sigma}W_3(M) \ .
\label{c1Pfaff}\eeq
It follows that ${\rm Pfaff}\otimes{\cal L}_B$ is trivial over $LM$ if
\beq
\phi^*\Bigl([H]\Bigr)=W_3(M) \ .
\label{Pfafftrivialcond}\eeq
The main consequence of this condition is that the D-brane worldvolume
$\phi:M\to X$ is {\it not} necessarily a spin$^c$ manifold
anymore. In particular, it no longer obeys Poincar\'e duality. Note
that in light of the discussion of the previous
subsection, the requirement (\ref{Pfafftrivialcond}) is also very
natural from a purely mathematical standpoint, as both sides of this
equation represent gerbes over $M$. The loss of the
spin$^c$ structure requires a modification of the K-theory groups used
to classify the D-brane charges.

Because of the condition (\ref{Pfafftrivialcond}), D-brane charge
should now live in an appropriate {\it twisted} version of
K-theory. Just as ordinary cohomology can be twisted by real line
bundles (i.e. elements of $\HH^2(X,\zed_2)$), so too can K-theory be
twisted by gerbes (i.e. elements of $\HH^3(X,\zed)$). The appropriate
receptacle for D-brane charge is then the K-theory of spacetime $X$
twisted by the characteristic class $[H]\in\HH^3(X,\zed)$ of the
$B$-field~\cite{Witten1}. A precise definition of this group has been
given in~\cite{KouwMathai1} based on algebraic
K-theory~\cite{WeggeOlsen}. Instead of using the $C^*$-algebra $C(X)$
of functions on spacetime, one uses the ring of sections
$\Gamma(X,{\cal E}_{[H]})$, where ${\cal E}_{[H]}\to X$ is the unique
locally trivial $PU({\cal H})$-bundle with Dixmier-Douady invariant
$[H]$. Thus the twisted K-theory groups are defined as
\beq
\K^*\Bigl(X\,,\,[H]\Bigr)=\K_*\Bigl(\Gamma(X,{\cal E}_{[H]})\Bigr) \ .
\label{twistedKtheory}\eeq
This group coincides with the K-theory of bundle gerbes obtained
  through the standard Grothendieck construction~\cite{BCMMS}. When
  $[H]=0$, the infinite rank bundle ${\cal E}_{[H]=0}=X\times{\cal
  K}({\cal H})$ is trivial and $\Gamma(X,{\cal
  E}_{[H]=0})=C(X)\otimes{\cal K}({\cal H})$. Then, by Morita
  equivalence, the K-theory group (\ref{twistedKtheory}) reduces to
  the usual one $\K^*(X,[H]=0)=\K_*(C(X)\otimes{\cal K}({\cal
  H}))\cong\K^*(X)$. When $[H]\neq0$ is a torsion class, the
  definition (\ref{twistedKtheory}) reproduces the standard definition
  in terms of the K-theory of Azumaya algebras~\cite{Witten1,Kapustin}
  which leads to the standard brane constructions described in section~2.3. The
  situations in which $[H]$ is a not a torsion class require, as in
  section~3.1, an infinite number of unstable D-branes in processes
  involving tachyon condensation~\cite{Witten2}.

\newsubsection{D-Brane Charge Lives in Twisted K-Homology}

The proposal (\ref{twistedKtheory}) for the twisted
K-theory groups fits in well with our current understanding of
D-branes in curved spaces. However, it does not demonstrate very well
the physical significance of the groups $\K^*(X,[H])$. A big quest at
present in this field is therefore to provide a clear physical
foundation to support the general claim that $\K^*(X,[H])$ is the
correct D-brane charge group to use in the presence of topologically
non-trivial $B$-field backgrounds. This problem has been addressed
within different contexts in~\cite{MMS}.

Following what we did in the previous sections, we can alternatively seek a
clearer geometric picture of twisted K-theory. At the functional
analytic level, this is relatively straightforward to do using twisted
Fredholm modules~\cite{MathSing}. The definition follows the same
pattern as before, except that now the algebra $C(X)$ is replaced with
$\Gamma(X,{\cal E}_{[H]})$ and one requires a $*$-homomorphism
\beq
\Phi_{[H]}\,:\,\Gamma(X,{\cal E}_{[H]})~\longrightarrow~
{\cal B}({\cal H}) \ .
\label{PhiH}\eeq
The corresponding equivalence classes generate the analytic K-homology
$\K_*(X,[H])$ of $X$ twisted by the characteristic class $[H]$ of the
$B$-field.

At the geometric level, we propose to interpret $\K_*(X,[H])$ as the
group spanned by certain triples~\cite{STWunpub}
\beq
k\Bigl(X\,,\,[H]\Bigr)={\rm Span}^{~}_\zeds\left\{(M\,,\,\phi\,,\,x)~\left|~
{\begin{array}{l}
M~{\rm closed,}~{\rm oriented}\\\phi\,:\,M~\longrightarrow~X\\
\phi^*\Bigl([H]\Bigr)=W_3(M)\\
x\in\K^*(M)\end{array}}\right.\right\} \ .
\label{kXHtriples}\eeq
Note that this definition differs from (\ref{kXtriples}) in the
weakening of the spin$^c$ condition, appropriate to a topologically
trivial background, to an orientability requirement ($w_1(M)=0$) and
the incorporation of the anomaly cancellation constraint
(\ref{Pfafftrivialcond}). It is clear that any D-brane in a
topologically non-trivial $B$-field defines an element of the abelian
group (\ref{kXHtriples}), and also of the twisted K-homology as
defined above. The definition (\ref{kXHtriples}) is then
the working proposal for twisted topological K-homology. The key point
which should make this the correct definition is as follows. The idea
is based on the fact that a vector bundle is orientable with respect
to complex K-theory if and only if it has a spin$^c$ structure. This
implies that the degree three Stiefel-Whitney class $W_3$ acts as an
``orientation gerbe'' for K-theory, so that for any space, K-theory
twisted by $W_3$ should give a version of Poincar\'e duality
appropriate to this setting. Then the K-theory class $x\in\K^*(M)$ has
a dual in $\K_*(M,W_3(M))$. The condition (\ref{Pfafftrivialcond})
then ensures that this will push-forward to a class in the twisted
K-homology $\K_*(X,[H])$ of spacetime $X$. The proof that this is
really the case should follow a method analogous to that in the
previous section,\footnote{\baselineskip=12pt There is a slight caveat
  though. Setting $[H]=0$ does not quite reduce (\ref{kXHtriples}) to
  its untwisted version~(\ref{kXtriples}), because in the latter case
  we also require each worldvolume $M$ to carry a specific, fixed choice of
  spin$^c$-structure. Presumably this problem can be solved by placing
  appropriate quotient conditions on~(\ref{kXHtriples}).} and at the
same time this will presumably shed light
on some physical properties of D-branes on curved manifolds. In
particular, using the recent construction~\cite{BCMMS} of the Chern
character $\ch_{[H]}$ in twisted K-theory, one should now be able to
rigorously derive a twisted version of the D-brane charge formula
(\ref{chargecohclass}).

\subsection*{Acknowledgments}

The author is grateful to R.~Minasian, P.~Seidel, P.~Turner, A.~Wassermann and
S.~Willerton for interesting discussions and remarks, and to K.~Hori
and P.~Turner for helpful comments on the manuscript. This work was
supported in part by an Advanced Fellowship from the Particle Physics
and Astronomy Research Council~(U.K.).

\setcounter{section}{0}

\end{document}